\documentclass{elsart}

\usepackage{epsfig,epsf}
\usepackage{graphicx}
\usepackage{epstopdf}
\newcommand{\beq}{\begin{equation}}
\newcommand{\eeq}{\end{equation}}

\def\fig#1{{Fig.~\ref{#1}}}

\newcommand{\as}{\alpha_s}
\newcommand{\un}{\underline}

\newtheorem{theorem}{Lecture~\#\!\!}
\newcommand{\bet}{\begin{theorem}}
\newcommand{\eet}{\end{theorem}}

\usepackage{amssymb}

\begin{document}

\begin{frontmatter}



\title{High energy pA collisions in the Color Glass Condensate approach}


\author{Kirill Tuchin}

\address{Department of Physics and Astronomy, Iowa State University, Ames, IA 50011}
\address{RIKEN BNL Research Center,
Upton, NY 11973-5000}


\begin{abstract}
We present a brief review of phenomenological applications of the gluon saturation approach to the proton-nucleus collisions at high energies. 

\end{abstract}

\end{frontmatter}

\section{Introduction: gluon saturation and structure functions}
\label{sec:intr}

At very high energies corresponding to very small values of Bjorken
$x$ variable the density of partons in the hadronic and nuclear wave
functions is believed to become very large reaching the {\it
saturation}  limit (see \cite{Iancu:2003xm,Jalilian-Marian:2005jf} and references therein). In the saturation regime  the
growth of partonic structure functions with energy slows down
sufficiently to unitarize the total hadronic cross sections. The
gluonic fields in the saturated hadronic or nuclear wave function are
very strong \cite{Iancu:2003xm,Jalilian-Marian:2005jf}. A transition to the saturation region
can be characterized by the {\it saturation scale} $Q_s^2 (s)$, which
is related to the typical two dimensional density of the partons'
color charge in the infinite momentum frame of the hadronic or nuclear
wave function. The saturation scale $Q_s^2 (s)$ is
an increasing function of energy $s$ and of the atomic number of the
nucleus A. At
high enough energies or for sufficiently large nuclei the saturation
scale becomes much larger than $\Lambda_{QCD}^2$ allowing for
perturbative description of the scattering process at hand. The presence of intrinsic large momentum scale $Q_s$
justifies the use of perturbative QCD expansion even for such a
traditionally non-perturbative observables as total hadronic cross
sections.
There has been a lot of activity devoted to calculating
hadronic and nuclear structure functions in the saturation regime. The
original calculation of quark and gluon distribution functions
including multiple rescatterings without QCD evolution in a large
nucleus was performed in \cite{Mue}. The resulting Glauber-Mueller
formula provided us with expressions for the partonic structure
functions which reach saturation at small $Q^2$. McLerran and
Venugopalan has argued  that the large density of gluons
in the partonic wave functions at high energy allows one to
approximate the gluon field of a large hadron or nucleus by a
classical solution of the Yang-Mills equations. The resulting gluonic
structure functions has been shown to be equivalent to the
Glauber-Mueller approach \cite{Iancu:2003xm,Jalilian-Marian:2005jf}. 

To include quantum QCD evolution in
this quasi-classical expression for the structure functions one has 
 to resum  the multiple BFKL pomeron
 exchanges.  The evolution equation resumming leading
logarithms of energy ($\as \ln s$) and the multiple pomeron exchanges
was written in \cite{yurieq} using the dipole model of
\cite{dip} and independently in \cite{bal} using the effective high
energy lagrangian approach. The equation was written for forward scattering amplitude $N({\underline
r},{\underline b} , Y)$ of a quark-antiquark dipole with transverse size $\underline r$ at impact
parameter $\underline b$ with rapidity $Y$ scattering on a target hadron or
nucleus, which, in turn can yield us the $F_2$ structure function of
the target which is measured in DIS experiments.

The evolution equation for $N$ closes only in the large-$N_c$ limit of
QCD and provides the basis for all phenomenological applications. It describes the onset of the gluon saturation for small $x$ and/or larger $A$.  The most striking consequence of the gluon saturation is scaling of the total DIS cross section at small $x$ with variable $Q^2/Q^2_s(x,A)$, the so-called geometric scaling, observed in DIS on both proton \cite{Stasto:2000er} and nuclear targets\cite{Armesto:2004ud}.

Several other observables can be calculated in the framework of the
saturation approach to hadronic and nuclear collisions.
In this paper we review applications of the saturation approach for calculation of various inclusive observables in pA collisions.

\section{Inclusive gluon production}

Single inclusive gluon production in the quasi-classical approximation has been derived in \cite{Kovchegov:1998bi}.  The inclusive cross  section for the scattering of a dipole of transverse $\underline x_{01}$ on the target reads 
$$
\frac{d {\hat \sigma}^{q{\bar q}A}_{incl}}{d^2 k \, dy}({\underline x}_{01}) 
=  \frac{\as C_F}{\pi^2}    \int  d^2 b  
d^2 z_1  d^2 z_2  e^{- i {\underline k} \cdot ({\underline z}_1 -
{\underline z}_2)}  \sum_{i,j=0}^1 (-1)^{i+j}
\frac{{\underline z}_1- {\underline x}_i}{|{\underline z}_1-
{\underline x}_i|^2} \cdot
\frac{{\underline z}_2- {\underline x}_j}{|{\underline z}_2- {\underline
x}_j|^2} 
$$
\beq\label{qcincl}
\times \frac{1}{(2 \pi)^2}\left( e^{- ({\underline x}_i- {\underline
x}_j)^2 Q_{0s}^2 /4} - e^{- ({\underline z}_1- {\underline x}_j)^2
Q_{0s}^2 /4} - e^{- ({\underline z}_2- {\underline x}_i)^2 Q_{0s}^2
/4} + e^{- ({\underline z}_1- {\underline z}_2)^2 Q_{0s}^2 /4} \right)
\eeq
This formula sums up the multiple rescatterings of the $q\bar q$ pair and the produced gluon on the nucleons in the target nucleus in the $A_+=0$ light cone gauge ($+$ is the direction of motion of the incident color dipole). Inclusion of quantum evolution amounts to resuming all possible real and virtual gluon emissions in addition to the emission of the measured gluon. In spite of the explicit breaking of factorization for individual diagrams in any known gauge the sum over all diagrams can be written in a simple $k_T$-factorized form \cite{Kovchegov:2001sc}
\beq
\frac{d\sigma^{pA}}{d^2k dy} =\frac{2\as}{C_F}\frac{1}{\underline k^2}\int d^2q\phi_p(\underline q)\phi_A(\underline k-\underline q)\,,
\eeq
where the unintegrated gluon distribution function is defined as
\beq
\phi(x,\un k^2)=\frac{C_F}{\as(2\pi)^3}\int d^2b d^2r e^{-i\un k\cdot \un r}\nabla^2_rN_G(\un r, \un b, y=\ln(1/x))\,,
\eeq
with $N_G(\un r, \un b, y)$ the forward amplitude of a \emph{gluon} dipole of transverse size $\un r$ at impact parameter $\un b $ and rapidity $y$ scattering on a nucleus. If the forward scattering amplitude of $q\bar q$ dipole is found from the evolution equation, then the gluon dipole scattering amplitude can be calculated from 
\beq
N_G(\un r, \un b, y)=2N(\un r, \un b, y)-N^2(\un r, \un b,y)\,.
\eeq
Let us emphasize that in this approach we consider gluon saturation effects only in nucleus, while treating proton as a dilute object.  Analysis of the inclusive gluon production has been also performed in \cite{Braun:2000bh,Dumitru:2001ux,Blaizot:2004wu}.

\begin{figure}[ht]
   \begin{tabular}{cc}
        \includegraphics[width=6cm]{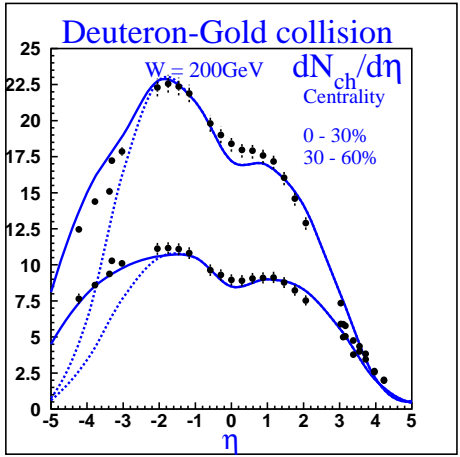}&
        \includegraphics[width=7cm]{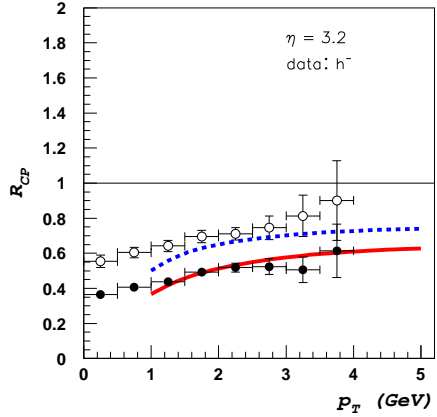}\\
        (a) &(b)
\end{tabular}
\caption{(a) Hadron multiplicity in d-Au collisions\cite{Kharzeev:2000ph,Kharzeev:2001gp} , (b) nuclear modification factor as a function of the transverse momentum $p_T$ at pseudo-rapidity $\eta=3.2$\cite{Kharzeev:2004yx}.}
\label{dAmult}
\end{figure}

This  approach to the high energy pA collisions leads to a rather successful phenomenological applications which allow the direct comparison with the experimental data, see \fig{dAmult}. 
While the multiple rescatterings in the nucleus without quantum evolution lead to mere redistribution of the gluon's transverse  momentum towards higher values (Cronin effect), the effect of quantum evolution is to tame the growth of the scattering amplitude at higher energies/rapidities and/or for heavy nuclei. As the result the saturation approach predicted the onset of suppression of the nuclear modification factor in dAu collisions at forward rapidities \cite{Kharzeev:2002pc,Kharzeev:2003wz,Baier:2003hr,Blaizot:2004wu,Iancu:2004bx}, see \fig{dAmult}.

\begin{figure}[ht]
   \begin{tabular}{cc}
        \includegraphics[width=6cm]{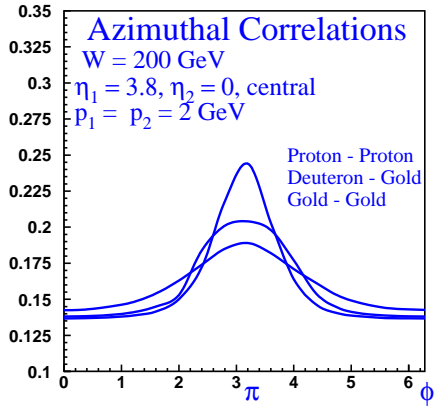}&
        \includegraphics[width=8.4cm]{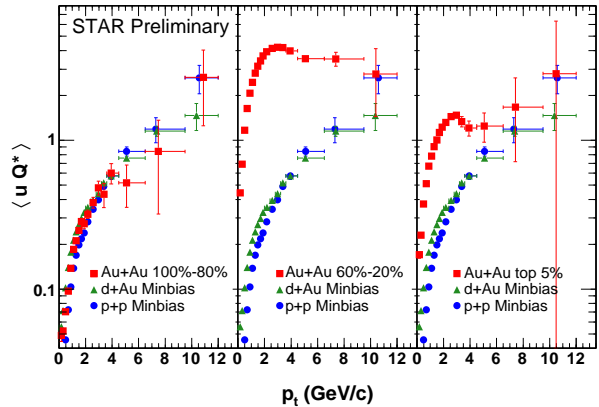}\\
        (a) &(b)
\end{tabular}
\caption{(a) Azimuthal correlations between forward and backward hadrons, (b) scaled elliptic flow variable $v_2\cdot M$ as a function of $p_T$ \cite{Tang:2004vc}.  $M$ is multiplicity.}
\label{fig:corr}
\end{figure}

It was suggested already in the pioneering papers on the gluon saturation that the particle correlations in the saturation regime must be significantly suppressed as compared to the low density regime (see \cite{Jalilian-Marian:2005jf}).  Indeed, unlike in hard processes where two jets are produced back-to-back, in saturation regime most gluons are produced with a semi-hard momentum of the order of $Q_s$ (``monojets"). From  the McLerran-Venugopalan model point of view, the emitted gluons correspond to the commuting classical non-Abelian fields. The double inclusive gluon production in the quasi-classical approximation has been addressed in \cite{Kovchegov:2002nf,Kovchegov:2002cd} where it was argued that the monojet correlations are responsible for generation of the azimuthal asymmetry in heavy ion collisions at large $p_T$. Quantum evolution effects where discussed in \cite{Kharzeev:2004bw} where it was predicted that suppression of correlations at forward rapidities is larger than at central ones and  that this suppression grows with the rapidity interval between the produced hadrons, see \fig{fig:corr}. Other possible signatures of the gluon saturation in the hadron correlation function were discussed in \cite{Baier:2005dv}.

Exact theoretical result for the double-gluon production case has been obtained in \cite{Jalilian-Marian:2004da}.  It turns out that the $k_T$-factorization fails. Instead a more complicated
factorization picture emerges.

\section{Heavy quark production}

Heavy quark production in hadronic collisions in high energy QCD is
one of the most interesting and difficult problems. It is
characterized by two hard scales: heavy quark mass $m$ and the
saturation scale $Q_s$.  The threshold for the invariant mass of the
quark $q$ and antiquark $\bar q$ production is $2m$. Therefore, if $m$
is much larger than the confinement scale $\Lambda_{QCD}$, it
guarantees that a non-perturbative long distance physics has little
impact on the quark production making
perturbative calculations possible  (for a review see
\cite{Brambilla:2004wf}).
 For all processes involving heavy
quarks with momentum transfer of the order of $Q_s^2\sim m^2$ large
saturation scale implies breakdown of the collinear factorization
approach. The factorization approach may be extended by allowing the
incoming partons to be off-mass-shell.  This results in conjectured
$k_T$-factorization.  Although the phenomenological
applications of the $k_T$-factorization approach seem to be
numerically reasonable at not very high energies
\cite{Fujii:2005vj} its theoretical status is not completely
justified. Like collinear factorization it is based on the leading
twist approximation. However, at sufficiently high energies, higher
twist contributions proportional to $(Q_s/m)^{2n}$ become important in
the kinematic region of small quark's transverse momentum, indicating
a breakdown of factorization approaches.

The fact that the saturation scale at high enough energies and for
large nuclei is large, $Q_s \gg \Lambda_{QCD}$, combined with the
observation that the typical transverse momentum of particles produced
in $pA$ scattering is of the order of that saturation scale, leads to
the conclusion that $Q_s$ sets the scale for the coupling constant,
making it small. This allows one to perform calculations for, say,
gluon production cross section in $pA$ collisions using the small
coupling approach \cite{Kovchegov:2001sc}. The same line of reasoning can be applied
to heavy quark production considered here: the saturation scale $Q_s$
is the important hard scale making the coupling weak even if the quark
mass $m$ was small. Having the quark mass $m$ as another large
momentum scale in the problem only strengthens the case for
applicability of perturbative approach. 

\begin{figure}[ht]
   \begin{tabular}{cc}
        \includegraphics[width=8cm]{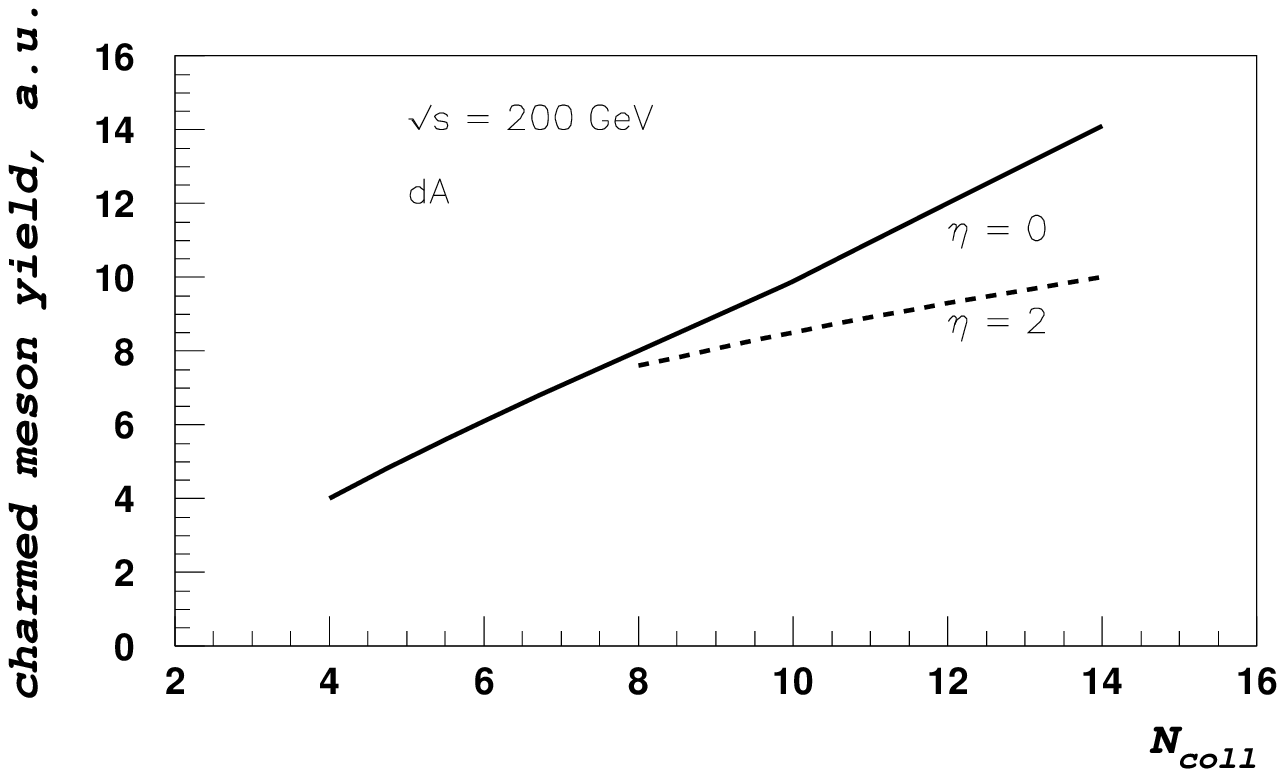}&
        \includegraphics[width=6cm]{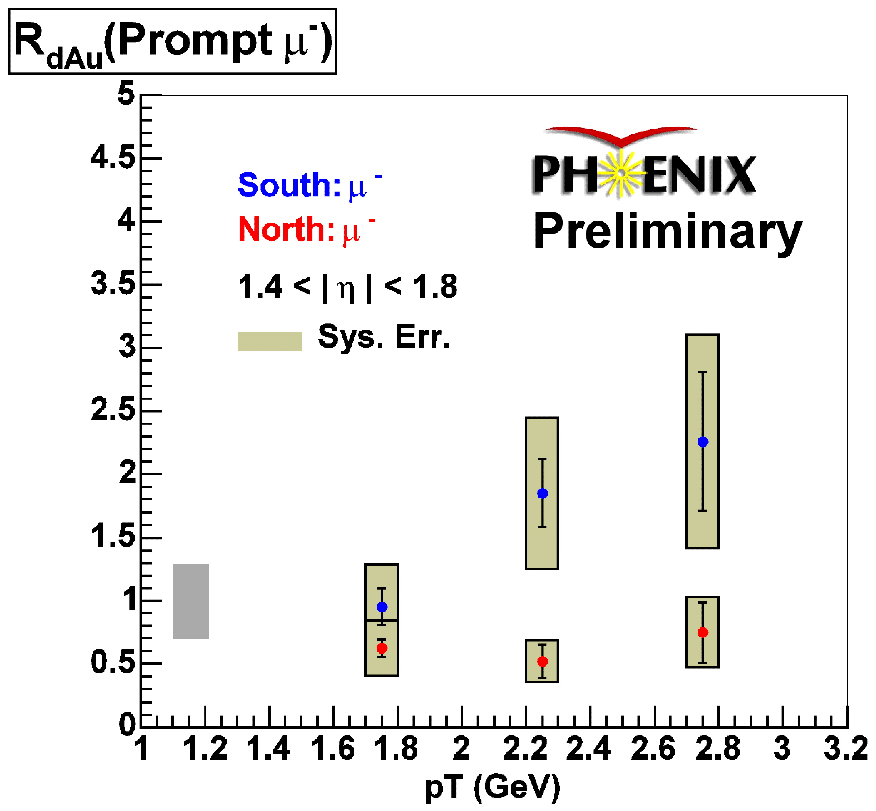}\\
        (a) &(b)
\end{tabular}
\caption{(a) Charmed meson yield as a function of $N_\mathrm{coll}$ \cite{Kharzeev:2003sk}, (b) Nuclear modification factor for muons associated with $D$-meson decays \cite{Averbeck}.}
\label{fig:quarks}
\end{figure}

Production of quark-antiquark
pairs in high energy proton-nucleus collisions and in DIS has been calculated in  
\cite{KopTar,Gelis:2003vh,Tuchin:2004rb,Blaizot:2004wv,Kovchegov:2006qn}.  
The results of the calculations in a model based on the $k_T$-factorization approach are displayed in \fig{fig:quarks}(a).  The heavy quark mass delays the onset of the gluon saturation effects to higher rapidities $\eta\simeq 2$ (for charm).  The recent experimental results for the nuclear modification factor for muons associated with $D$-meson decays are consistent with this prediction, see  \fig{fig:quarks}(b).

\section{Future perspectives}

In this admittedly brief review article we discussed how the effect of gluon saturation can be taken into account in the framework of the perturbation theory in the case of gluon and quark spectra, multiplicities and correlations. These observables are in agreement with the recent experimental data. 
In addition to the aforementioned calculations of the gluon and quark spectra and multiplicities an extensive work has been done on computing the production of $J/\Psi$\cite{Kharzeev:2005zr}, valence quarks\cite{Dumitru:2002qt,Gelis:2001da}, prompt photons\cite{Jalilian-Marian:2005zw} and di-leptons\cite{Baier:2004tj,Jalilian-Marian:2004er,Betemps:2004xr}. These observables still require detailed experimental investigation. Although we observed the onset of the gluon saturation at RHIC, this effect will be far more dramatic at LHC. In fact, production of bulk of particles in AA and pA collisions at midrapidity at LHC is predicted to be governed by the saturation regime.
 Measurement of the energy dependence of the various physical quantities is an important test for the gluon saturation approach.


{\bf Acknowledgments.}
 The author would like to thank RIKEN, BNL and the U.S. Department of Energy (Contract
No. DE-AC02-98CH10886) for providing the facilities essential for the
completion of this work.

\end{document}